\documentclass[aps,twocolumn,amssymb,floatfix,amsmath,prb]{revtex4-1}
\usepackage{epsfig}
\usepackage{graphicx}%
\usepackage{epstopdf}
\usepackage{dcolumn}
\usepackage{bm}
\usepackage{ifthen}
\usepackage{booktabs}

\newcommand{\affstanford}{Stanford Synchrotron Radiation Lightsource, Stanford University, Menlo Park, CA 94025, USA}

\begin{document}

\author{I. Lorite}\email{lorite@physik.uni-leipzig.de}
\affiliation{Institut f\"ur Experimentelle Physik II, University
of Leipzig, Linn\'estra{\ss}e 5, D-04103 Leipzig, Germany}
\author{C. Zandalazini}
\affiliation{Institut f\"ur Experimentelle Physik II, University of Leipzig, Linn\'estra{\ss}e 5, D-04103 Leipzig, Germany}
\altaffiliation{On leave from: Laboratorio de F\'{i}sica del S\'{o}lido, Dpto. de F\'{i}sica, FCEyT, Universidad Nacional de
Tucum\'{a}n, Argentina}

\author{P. Esquinazi}
\affiliation{Institut f\"ur Experimentelle Physik II, University of Leipzig,
                  Linn\'estra{\ss}e 5, D-04103 Leipzig, Germany}

\author{D. Spemann}
\affiliation{Institut f\"ur Experimentelle Physik II, University of Leipzig,
                  Linn\'estra{\ss}e 5, D-04103 Leipzig, Germany}
\author{S. Friedl\"{a}nder}
       \affiliation{Institut f\"ur Experimentelle Physik II, University of Leipzig,
                    Linn\'estra{\ss}e 5, D-04103 Leipzig, Germany}
\author{A. P\"oppl}
       \affiliation{Institut f\"ur Experimentelle Physik II, University of Leipzig,
                    Linn\'estra{\ss}e 5, D-04103 Leipzig, Germany}

\author{T. Michalsky}
       \affiliation{Institut f\"ur Experimentelle Physik II, University of Leipzig,
                    Linn\'estra{\ss}e 5, D-04103 Leipzig, Germany}

\author{M. Grundmann}
       \affiliation{Institut f\"ur Experimentelle Physik II, University of Leipzig,
                    Linn\'estra{\ss}e 5, D-04103 Leipzig, Germany}

\author{J. Vogt}

       \affiliation{Institut f\"ur Experimentelle Physik II, University of Leipzig,
                    Linn\'estra{\ss}e 5, D-04103 Leipzig, Germany}

\author{J. Meijer}
       \affiliation{Institut f\"ur Experimentelle Physik II, University of Leipzig,
                    Linn\'estra{\ss}e 5, D-04103 Leipzig, Germany}

\author{S. P. Heluani}
       \affiliation{Laboratorio de F\'{i}sica del S\'{o}lido, Dpto. de F\'{i}sica,
                    FCEyT, Universidad Nacional de Tucum\'{a}n, Argentina.}
                    \author{H.~Ohldag}
\affiliation{\affstanford}
\author{W. A. Adeagbo}
       \affiliation{Institute of Physics, Martin Luther University Halle-Wittenberg,
                    Von-Seckendorff-Platz 1, 06120 Halle, Germany}%
\author{S. K. Nayak}
       \affiliation{Institute of Physics, Martin Luther University Halle-Wittenberg,
                    Von-Seckendorff-Platz 1, 06120 Halle, Germany}%
\author{W. Hergert}
       \affiliation{Institute of Physics, Martin Luther University Halle-Wittenberg,
                    Von-Seckendorff-Platz 1, 06120 Halle, Germany}%
\author{A. Ernst}
       \affiliation{Max Planck Institute of Microstructure Physics, Weinberg 2,
                    06120 Halle, Germany}
\author{M. Hoffmann}
       \affiliation{Max Planck Institute of Microstructure Physics, Weinberg 2,
                    06120 Halle, Germany}
                     \affiliation{Institute of Physics, Martin Luther University Halle-Wittenberg,
                    Von-Seckendorff-Platz 1, 06120 Halle, Germany}%

\title{Study of the Negative Magneto-Resistance of Single Proton-Implanted Lithium-Doped ZnO Microwires}

\begin{abstract}
The magneto-transport properties  of single proton-implanted ZnO and of Li(7\%)-doped  ZnO microwires have been studied. The as-grown microwires were highly insulating
and not  magnetic.
After proton implantation the Li(7\%) doped ZnO  microwires showed a non monotonous behavior of
the negative magneto-resistance (MR) at temperature above 150 K. This is in contrast to the monotonous NMR observed below 50 K
for proton-implanted ZnO.
The observed difference in the transport properties of the wires is related to the amount of
stable Zn vacancies created at the near surface region by the proton implantation and Li doping.
The magnetic field dependence of the resistance might be explained by the formation of a
magnetic/non magnetic heterostructure in the wire after proton implantation.

\end{abstract}

\maketitle


\section{Introduction}

A large number of reports indicates
that  vacancies and/or the doping with nonmagnetic ions play a
main role in triggering defect-induced magnetism (DIM)  in ZnO\cite{Bhos08,kha09,ogale10,sto10,vol10,dim13}.
In particular, It has been shown experimentally and
theoretically that in ZnO, Zn vacancies (V$_{\text{Zn}}$)
are the main defect that, for a concentration of $\sim 5~$\%, can trigger magnetic order at
temperatures T $\geq  300~$K.\cite{kha09,hau11,awa12}
It has been suggested that room temperature magnetic order is better
stabilized by hole doping ZnO,  arguing that
holes mediate the long range coupling between localized magnetic moments.\cite{peng09}

Previously, lithium, as an element that can be easily incorporated into ZnO,
was already used to produce  p-type ZnO.\cite{lee11,fan13} However, the  efficiency of
p-doping with Li is generally limited by the formation of
compensating interstitials,  the spontaneous formation of
opposite-charged  defects (e.g., cations or vacancies),
which pin or may even decrease the Fermi energy. In this case there are not
enough charge carriers generated at the working
temperature, increasing strongly the resistivity of the material.
The work in  Ref.~\onlinecite{awa12}
 characterized the existing defects in
ZnO nanoparticles doped with different concentrations of Li and
concluded that the observed magnetic order at room temperature is related
to Li and Zn defects, in particular  Li influences the formation
and stabilization of Zn vacancies, generating  the
predicted $p$-type ferromagnetism.\cite{peng09}

 In our previous report we presented evidence of the magnetic order of the ZLH wires only after H$^+$ implantation for a minimum Li concentration of 3\%.\cite{lor15}
The observed magnetism is due to the spin polarization of the O-2p band. This spin polarization occurs due to the proximity to the magnetic moments at the Zinc vacancies.
These are produced during H$^+$ implantation and stabilized thanks to the Li doping \cite{lor15}. Therefore, the Li concentration fixes the amount of V$_{\rm Zn}$  large enough to produce magnetic ordser at room temperature.

We note that in general the main property used in literature to prove the existence of DIM
in ZnO is the bulk magnetization, even in small oxide structures
as  ZnO nanorods \cite{pan10} or  ZnO nanoparticles \cite{li13}. However,
 in spite of the importance for the future
miniaturization of spintronic devices with high-temperature
magnetic order and low resistivities, evidence of ferromagnetic behavior at room temperature
in a single micro- or nanostructure of ZnO has not been  reported in literature  yet.
The present work reports on the magneto-transport properties of low resistive Li-doped ZnO
microwires after proton implantation. We believe that the observed
transport properties are of interest for the development of spintronic devices based on these materials.

\section{Experimental}
ZnO and Li-doped ZnO microwires were prepared by a carbothermal process as explained
elsewhere\cite{lor15,lor14}. The studied wires had a diameter between $\simeq 0.5~\mu$m and $\simeq
10~\mu$m and a length of some $\simeq 100~\mu$m.
The percentage of Li was chosen following
previous reports \cite{cha07,chaw09,ran12}.
The H$^+$ implantation of the ZnO microwires was performed in a
remote hydrogen DC-plasma chamber in parallel-plate configuration during 1 hour
at a current of 60~$\mu$A.\cite{lor14,kha12} Assuming a displacement energy of 18.5~eV and 40~eV for
Zn and O in pure ZnO lattice \cite{Look99} SRIM simulation indicates that it is
possible to create both V$_{\text{Zn}}$  and
O-vacancies (V$_{\text{O}}$) within the first 10~nm from the surface of ZnO
due to the low energy of H$^{+}$ implantation used in this work.

The pure ZnO and ZnO:Li(7\%) as well as H$^+$-implanted ZnO and  ZnO:Li(7\%)
microwires are labeled as ZnO, ZL, ZH and ZLH, respectively.
Single microwires were selected using an optical microscope and
fixed on a Si/Si$_3$N$_4$ substrate. The electrical contacts were done by
clenching gold wires with indium on the microwire providing ohmic
contacts at all measured temperatures. The magneto transport
measurements were performed in a He-cryostat with the possibility to apply a magnetic field
of 8~T  at a maximum temperature of $\simeq 250~$K.
The rotating sample holder allowed us measurements of
the magneto transport properties at fields applied perpendicular
and parallel to the electrical current direction, which is along
the  main axis of the wire.

Electron paramagnetic resonance (EPR) measurements were
performed with a BRUKER EMX Micro X-band spectrometer at
9.41$\,$GHz with an Oxford ESR 900 flow cryostat at temperatures
from $6 - 300\,$ K in the dark and under illumination with the full
spectrum of a  Xe-lamp. The sample volume was $\simeq 15\, $mm$^3$
loosely packed ZnO microwires, previous to Li doping, in a standard test tube. EPR
spectra were simulated by using the Easy Spin Matlab toolbox
\cite{sto06}.

\section{Results and discussion}

\begin{figure}
\centering{
\includegraphics[width=1.0\columnwidth]{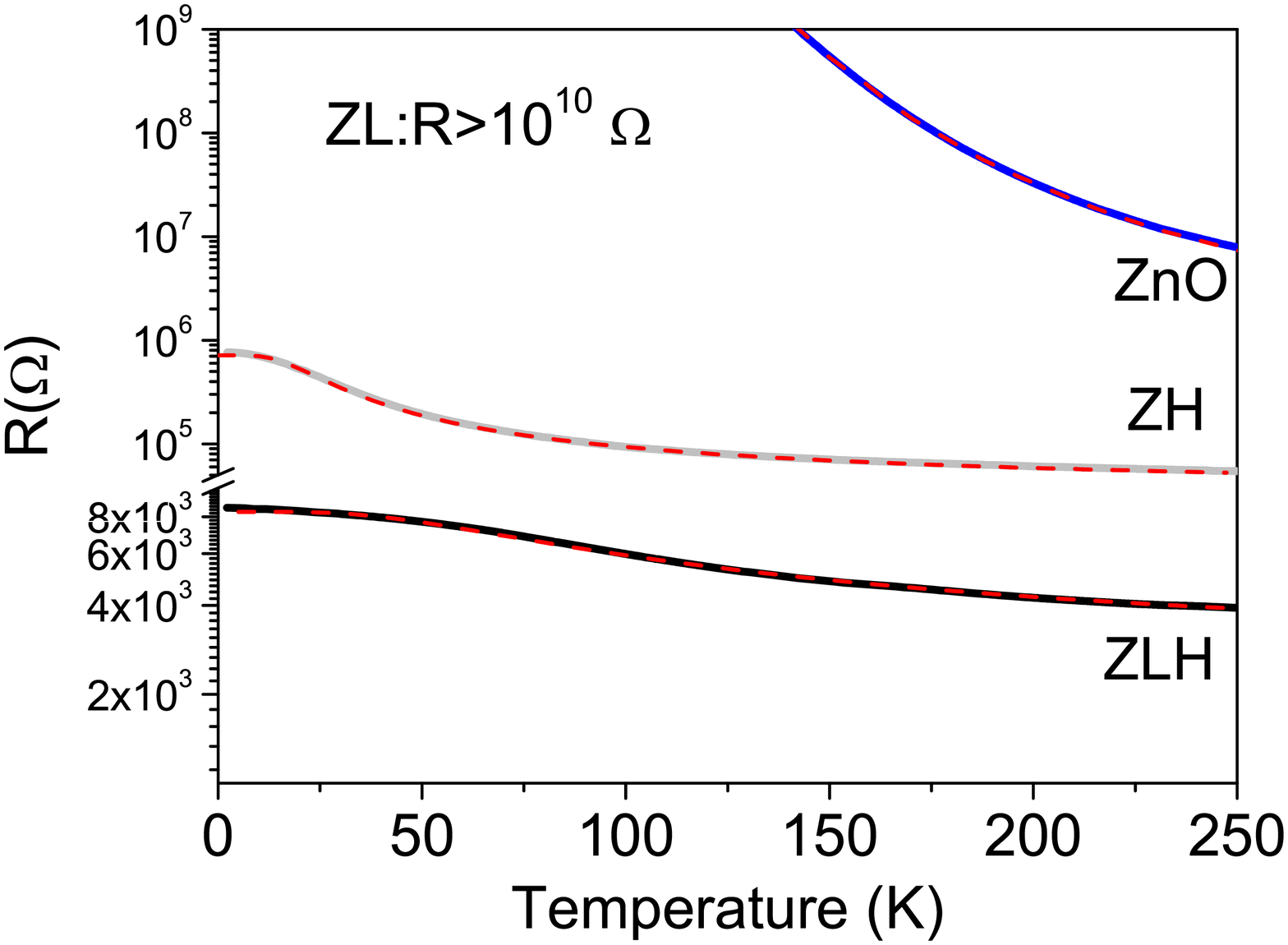}}
 \caption{Resistance of all measured ZnO, ZH, ZL and ZLH microwires
 of similar length and diameter ($\simeq 300~\times 10~\mu$m$^2$) as a function of temperature in a semi-logarithmic scale.
 Dashed red lines are the fittings of the experimental data to Eq.~(\ref{vrh}).}
 \label{RT}
 \end{figure}

Figure \ref{RT} shows the temperature dependence of the resistance
of different ZnO wires with similar geometry. Pure ZnO wire shows a semiconducting
like behavior with $E_g \simeq 0.28~$eV obtained from the fit to Eq.(1) ($R_s$ term).
After doping with Li (ZL), the
wire shows a highly insulating behavior having a resistance larger
than $10^{10}~\Omega$ (shown by an arrow in Fig.~\ref{RT}) at all temperatures. The
highly insulating behavior of ZL is related to the carrier compensation by the
complex formation of interstitial Li ions as donors with  Li ions
at Zn places as acceptors. The Zn vacancies act themselves as acceptors and
could provide a p-type character to the wire \cite{yi10}.

After H$^+$ implantation the wires show a large decrease in the resistance
of several orders of magnitude, see Fig.~\ref{RT}.
As  in the case of H$^+$ doped ZnO single crystals
\cite{kha12}, the temperature ($T$) dependence of the ZLH and ZH
microwires can be described using a simple parallel resistor
model, considering two different contributions, namely: one
 with the typical Arrhenius dependence for semiconductors
($R_s$) and an activation energy $E_g$. The second one is
given by a variable range hopping-like ($R_{VRH}$) mechanism that prevails
at lower $T$. It is reasonable to assume that the VRH mechanism occurs  mainly at the near
surface region of the wire, whereas the bulk retains its thermally activated semiconducting behaviour
with the corresponding  activation energy.

The equation for the total resistance based on the above described  model  can be written as:
\begin{eqnarray}
    R(T)&= &( R_s^{-1} + R_{VRH}^{-1} )^{-1} \nonumber\\
    &=& [ (R_1 \exp(E_g/2k_BT))^{-1} + \nonumber\\
    &+&(R_2 \exp(E_n/T)^p)^{-1}  ]^{-1}\,
    \label{vrh}
\end{eqnarray}
From the fits (dashed lines in Fig.~\ref{RT}) we obtain the
following values for the ZH (ZLH) microwires: $R_1 =
46~(3.5)~$k$\Omega$, $E_g = 13~(35)~$meV, $R_2 =
278~(3)~$k$\Omega$, and $E_{n} = 4~$meV and $p = 1/4$ for both.
The large variation of the E$_g$ from the undoped ZnO sample to
the H$^+$ treated ones is related to the extra impurity band
produced by the corresponding increase in the vacancies
concentration and the hydrogen doping.

According to the fits the main difference between the  ZH and ZLH wires is in the
prefactor on the VRH
part. This is because  the H$^+$ implantation affects mainly the near surface region, as expected from
the used implantation energy.

\begin{figure}
\centering{
\includegraphics[width=1.0\columnwidth]{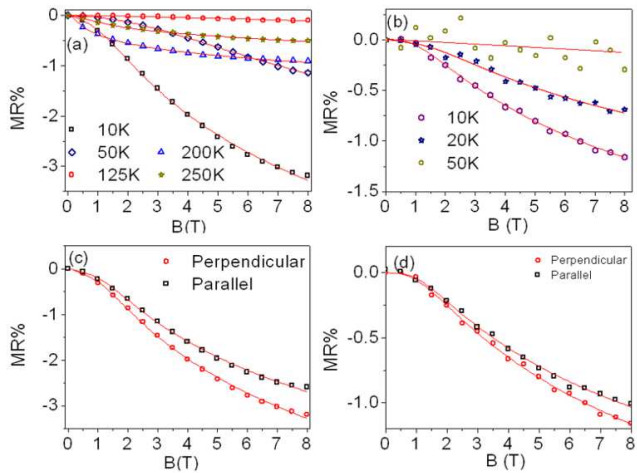}}
 \caption{Magnetoresistance in perpendicular configuration at different temperatures of the ZLH (a) and ZH (b) microwires. (c) and (d) figures show a comparison of the
 magnetoresistance at 10~K for parallel and perpendicular configuration of ZLH (c) and ZH (d). All the continuous line through the data are fits to Eq.~(\ref{mre}).}
 \label{MR}
 \end{figure}

\begin{figure}
\centering{
\includegraphics[width=1.\columnwidth]{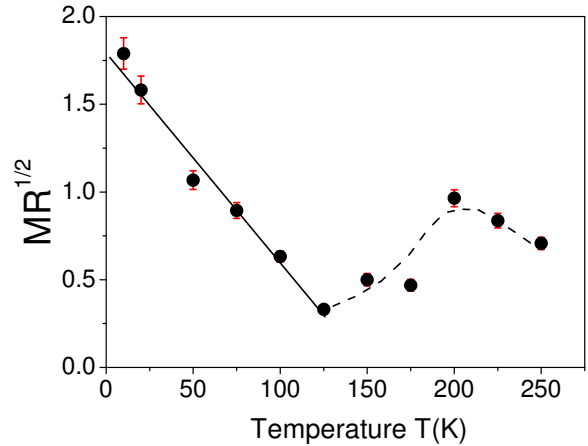}}
 \caption{Square root of the absolute value of the magnetoresistance at a magnetic field of 8~T  vs. temperature
 for a single ZLH microwire. The solid and the dashed lines are only a guide for the eye.}
 \label{MR12}
 \end{figure}

Figure \ref{MR} shows the
magnetoresistance (MR)  between 0~T and 8~T magnetic field
applied perpendicular to the current flow or $c$-axis of the ZLH (a) and ZH (b) microwires.
Figures \ref{MR}(c) and (d)  show the MR for field applied parallel and perpendicular to the $c$-axis of ZLH and ZH microwires
at different temperatures. The obtained results indicate: (1) For the
ZLH microwire a negative MR (NMR) of -3.2\% at 10~K (-0.5\% at 250~K) is measured
at 8~T applied field, see Fig.~\ref{MR}(a). (2) At the same applied field the ZH
microwire presents a lower NMR of -1\% at 10~K, being
negligible above 50~K, see Fig.~\ref{MR}(b). (3) For the ZLH wire the
NMR at a given fixed field is non-monotonous
in temperature, see Fig.~\ref{MR12}.
(4) A comparison of the NMR between fields applied parallel and normal
to the applied current (which has the same direction as
the main axis of the wires) at 10~K, see Figs.~\ref{MR}(c) and \ref{MR}(d),
indicates that the (absolute) MR is smaller  for
parallel fields than for fields applied normal to the current. The
observed variation with the angle is compatible with the
anisotropic MR, usually seen in ferromagnetic systems
\cite{handley00}.
We found that the square root of the (absolute) MR of the ZLH microwire decreases linearly with temperature to 120~K,
at higher temperatures the  MR increases again showing a
maximum at $T \sim 200~$K, see Fig.~\ref{MR12}.
The large and non-monotonous change of the MR with temperature is
related to different transport contributions. In particular the square temperature dependence
of the MR below 120~K appears to be related to a mechanism occurring at
the near surface region of the wire, where the VRH prevails.
We stress that a similar behavior of the MR was observed in H$^+$ doped
ZnO single crystals, but the minimum of the MR (at  fixed field) occurred at $\simeq 40~$K and the maximum at $\simeq 100~$K,\cite{kha12} instead of 125~K and 200~K as in our case, see Fig.~\ref{MR12}.
A possible origin of this effect is given below.

\begin{figure}
\centering{
\vspace{-1.0cm}
\includegraphics[width=1\columnwidth]{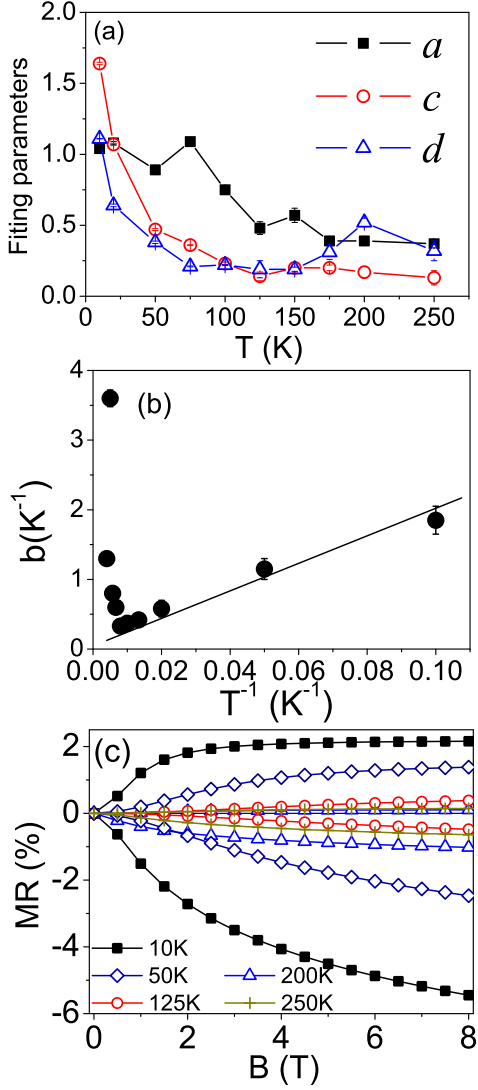}}
\vspace{-1.5cm}
 \caption{Coefficients $a,c,d$ (a) and $b$ (b) obtained from the fits to  Eq.~(\ref{mre}) of the data
 presented in Fig.~\ref{MR12} for the ZLH microwire. (c) The magnetoresistance calculated from the two contributions, the negative  (first term in  Eq.~(\ref{mre})) and positive one (second term in  Eq.~(\ref{mre})),
 using the parameters of the fits.}
 \label{coe}
 \end{figure}

We describe  the MR field dependence of ZLH wire with a semiempirical model proposed  by Khosla and Fischer
\cite{Kho70},  widely used for magnetic transition metals in the
past.
The model takes into account two field dependent
contributions. The positive
MR contribution with a quadratic field dependence at low fields and saturation
at high fields (second term in the r.h.s. of Eq.~(\ref{mre})) is due to two
conduction bands (usually $s$ and $d$) with different
conductivities. The negative MR one  (first term in the r.h.s. of
Eq.~(\ref{mre})) is attributed to a spin dependent scattering between
two sub-bands.
Strictly speaking it  does
not saturate  at large fields, but its absolute value increases
following a logarithmic field dependence. The semi-empirical formula is given by:

\begin{eqnarray}
%
 %
%
   \frac{\Delta R}{R(0)}&=& - a^2 \ln(1 + b^2B^2) + c^2B^2 / (1 + d^2B^2),
   \label{mre}
\end{eqnarray}
where $a, b, c$ and $d$ are free parameters that depend on the
carrier mobility, spin scattering amplitude, exchange integral
conductivity and the spin of the localized moments; $R(0)$ is the
resistance at zero field. The red lines of Fig.~\ref{MR} follow Eq.~(\ref{mre})
with the  coefficients shown in Fig.~\ref{coe}
obtained from the fits.

In Fig.~\ref{coe}(a) the coefficients $a, c, d$ and in Fig.~\ref{coe}(b) the coefficient $b$,
obtained from the fits of Eq.~(\ref{mre}) to the data of the ZLH wire, are shown. The parameters $a$ and $c$ decrease monotonously with the increase of
temperature up to  150~K.  Above such a temperature
they remain rather constant within the fitting error. The anomalous behaviour in the
MR is mainly reflected in the dependence of the fitting parameters $b$ and to some extent in  $d$, see Figs.~\ref{coe}(b) and (a). Following the original model \cite{Kho70} the parameter $b$ should be inversely proportional to the temperature. This dependence is indeed observed in the temperature range where the
VRH prevails, Fig.~\ref{coe}(b). However, this is not the case at higher temperatures.
A simulation of the negative and positive
contributions to the MR was done separately taking into account
the obtained parameters, see Fig.~\ref{coe}(c).
We observe that the positive MR contribution decreases monotonously with  temperature.
However, the negative MR present the expected non monotonous behavior in agreement with Fig.\ref{MR12} and seems to be affected by two different contributions. The first one is active within the temperature range where the VRH mechanism overwhelms, i.e.  below 150~K as the temperature dependence of the resistance suggests,  see Fig.~\ref{RT}. The second one takes place once the carriers of the semiconductor are thermally activated, i.e. above 150 K.

\begin{figure}
\centering{
    \includegraphics{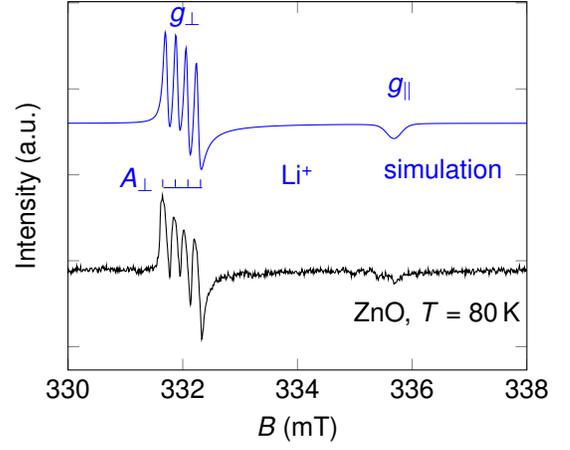}}
    \caption{ Experimental and simulated EPR spectrum  of a $\,^7$Li$^x_\text{Zn}$ center
    found in the pure ZnO sample measured at 80$\,$K with Xe-light illumination.}
    \label{fig:epr}
\end{figure}

Since for the ZnO microwire and after implantation of
a comparable H$^+$ dose (ZH) there is a vanishing of the NMR at T$ \>$ 50~K, see Fig.~\ref{MR}(b),
it appears reasonable to assume that the density of V$_{\text{Zn}}$ after implantation
is smaller than that for ZLH.
In order to clarify the origin for a possible decrease in the V$_{\text{Zn}}$ concentration we
have studied the agglomerate of ZnO microwires using EPR under UV-light to check for
the possible existence of Li impurities, which can stabilized a small amount of V$_{\text{Zn}}$.
Figure~\ref{fig:epr} shows, as an example, the signal obtained
for ZnO recorded at 80$\,$K under illumination. This temperature was
selected due to the larger signal to noise ratio taking into account that the
EPR signal decreases with temperature.\cite{bar10}
The intense four line spectrum at 332$\,$mT is
attributed to a $\,^7$Li$^x_\text{Zn}$ center with $S=\frac{1}{2}$ and $\, ^{\text{Li}}I=\frac{3}{2}$ which can be described by a spin Hamiltonian with axial symmetry
\begin{eqnarray}
    \hat{{H}}& =& \beta \left( g_\parallel \hat{S}_z {B}_z + g_\perp \left[ \hat{S}_xB_x +\hat{S}_yB_y \right] \right) +\\ \nonumber
    &&A_\parallel \hat{S}_z\, ^{\text{Li}}\hat{I}_z +
A_\perp \left( \hat{S}_x\, ^{\text{Li}}\hat{I}_x + \hat{S}_y\, ^{\text{Li}}\hat{I}_y \right)
    \label{eqn:hamiltonian}
\end{eqnarray}
where $\beta$ is the Bohr magneton, $g_\perp,\,g_\parallel,\,A_\perp$ and $A_\parallel$ are the principal values of the Zeeman splitting tensor $\hat{g}$ and of the Li hf coupling tensor $\hat{A}$. The electron and nuclear spin operator components are denoted as $\hat{S}_i$ and $\, ^{\text{Li}}\hat{I}_i$, respectively with $i=x,y,z$. The spin Hamiltonian parameters of the $\,^7$Li$^x_\text{Zn}$ center, $g_\parallel = 2.0028,\,g_\perp = 2.0251,\,A_\parallel =0.1\,$MHz, and $A_\perp=5.1\,$MHz as evaluated from the EPR spectrum shown in Fig.~\ref{fig:epr} are in good agreement with previously reported data of occupied Li acceptors in ZnO, where the Li$^+$ substitutes Zn$^{2+}$.\cite{sch68}

The obtained EPR signal reveals, therefore the existence of  Li
impurities  in nominally pure ZnO. Due to technical reasons is not
possible, however, to use the absolute magnitude of the EPR signal
to obtain a value of the Li content in the samples. These Li
impurities will help to stabilize V$_{\text{Zn}}$ after H$^+$
implantation. Since the concentration is small the stabilized
V$_{\text{Zn}}$ concentration will be also small. This
experimental result explains why  magnetic order after implanting
H$^+$ in ZH microwires is not observed, although a small NMR was
observer at low temperatures.

As previously mentioned, two different contribution must take place in the observed field dependence of resistance.
To account for the negative contribution within the VRH range, the model of static magnetic polaron \cite{coe05,lor14}
may be used. According to this model, shallow donors form bound magnetic polarons which overlap to create a spin-splitting in the band. At higher temperatures, this contribution must be reduced and so the NMR, as observed. It has to be noted that for a small density of magnetic defects the VRH vanishes at the temperature of 50 K, as presented for ZH wire.
However, this is not the case for ZLH, where the V$_{\text{Zn}}$ density is enough to produce magnetic order within the 10 nm near surface region meanwhile the rest of the wire remains non magnetic. The magnetic order only in the near surface region will produce, therefore a DIM/non-DIM heterostructure.
To explain the observed anomalous behaviour in the NMR of ZLH, we have to considere the interface between magnetic and non magnetic regions. With the rise of temperature, it is produced an increase of carrier concentration. This increase of carriers may contribute to enhance the DIM/non-DIM interfacial magnetism \cite{zhiy13} at a certain temperature, i.e. 150 k - 200 K,  and so the spin polarization resulting in more negative value of MR, see Fig.~\ref{MR12}.
This effect could help in the understanding and development of magnetic
heterostructure for spintronic devices such as spin valves\cite{tang15} through magnetic semiconductors.

\section{Conclusions}
In summary,  the negative magnetoresistance of single H$^+$- and Li-doped ZnO
microwires as well as the anisotropic magnetoresistance provide
further evidence for the existence of magnetic order. The behavior
of the magnetoresistance is  non monotonous with
temperature. This behavior could be related to an enhancement  of the
spin polarization occurring at the interface between magnetic and non-magnetic regions of the
proton implanted Li-doped microwires due to the increase of carrier concentration with temperature.


This work  was partially  supported  by CIUNT under Grants 26/E439
and 26/E478, by ANPCyT-PICTR 35682, and by the Collaborative
Research Center SFB~762 ``Functionality of Oxide Interfaces''. We
are grateful for the support within the DFG priority program
SPP~1601 ``New Frontiers in Sensitivity for EPR Spectroscopy''.


%

\end{document}